\documentclass[conference]{IEEEtran}

\usepackage{amsmath,mathtools}
\usepackage{amssymb}
\usepackage{pifont}
\usepackage{tikz}
\usetikzlibrary{positioning, arrows.meta, fit, shapes}

\usepackage{graphicx}
\usepackage{amsmath}
\usepackage{algorithm}
\usepackage{subfig}
\usepackage{algpseudocode}
\usepackage{boxedminipage}
\usepackage{fancyhdr}
\usepackage{multirow}

\usepackage{array}
\usepackage{textcomp}
\usepackage{stfloats}
\usepackage{csquotes} 
\usepackage{url}
\usepackage{verbatim}
\usepackage{cite}

\usepackage{eqnarray}
\def\BibTeX{{\rm B\kern-.05em{\sc i\kern-.025em b}\kern-.08em
    T\kern-.1667em\lower.7ex\hbox{E}\kern-.125emX}}
\usepackage{balance}
\usepackage{url}
\usepackage[top=0.75in, bottom=1in, left=0.75in, right=0.75in]{geometry}

\begin{document}

\title{Knowledge-Defined and Twin-Assisted Network Management for 6G}

\author{
    \IEEEauthorblockN{Tuğçe BILEN\IEEEauthorrefmark{1} and Mehmet Ozdem\IEEEauthorrefmark{2}}
    
    \IEEEauthorblockA{\IEEEauthorrefmark{1}Department of Artificial Intelligence and Data Engineering\\Faculty of Computer and Informatics\\
    Istanbul Technical University, Istanbul, Turkey\\}
    \IEEEauthorblockA{\IEEEauthorrefmark{2}Turk Telekom, Istanbul,Turkey \\
    Email: bilent@itu.edu.tr, mehmet.ozdem@turktelekom.com.tr}
}

\maketitle

\begin{abstract}
The increasing complexity, dynamism, and heterogeneity of 6G networks demand management systems that can reason proactively and generalize beyond pre-defined cases. In this paper, we propose a modular, knowledge-defined architecture that integrates Digital Twin models with semantic reasoning and zero-shot learning to enable autonomous decision-making for previously unseen network scenarios. Real-time data streams are used to maintain synchronized virtual replicas of the physical network, which also forecast short-term state transitions. These predictions feed into a knowledge plane that constructs and updates a graph-based abstraction of the network, enabling context-aware intent generation via graph neural reasoning. To ensure adaptability without retraining, the management plane performs zero-shot policy matching by semantically embedding candidate intents and selecting suitable pre-learned actions. The selected decisions are translated and enforced through the control plane, while a closed-loop feedback mechanism continuously refines predictions, knowledge, and policies over time. Simulation results confirm that the proposed framework observes notable improvements in policy response time, SLA compliance rate, and intent matching accuracy.
\end{abstract}


\thispagestyle{fancy}

\pagestyle{fancy}
\fancyhf{}
\fancyhead[C]{\scriptsize Accepted by Workshop on Data Driven and AI-Enabled Digital Twin Networks and Applications (TwinNetApp) in IEEE Global Communications Conference (Globecom), ©2025 IEEE}
\renewcommand{\headrulewidth}{0pt}

\section{Introduction}
The sixth generation (6G) of wireless networks is envisioned to push the boundaries of connectivity by enabling intelligent, autonomous, and context-aware communication systems \cite{BILEN201824}. Building on the foundations of 5G, 6G aims to support ultra-reliable low-latency communications (URLLC), massive machine-type communications (mMTC), and extreme data rates, while also introducing advanced capabilities such as integrated sensing, native artificial intelligence, and full network automation \cite{BILEN2020101133}, \cite{10463676}. These advancements are expected to unlock a wide range of applications where the network is an active decision-making entity.

Despite this ambitious vision, managing the complexity of 6G networks presents a number of critical challenges \cite{10078095}, \cite{9520342}. As networks become more dynamic and heterogeneous, it becomes increasingly difficult to anticipate every possible situation that may arise. Traditional rule-based or even supervised learning-based network management techniques rely heavily on pre-defined scenarios and labeled data, making them inadequate for dealing with previously unseen conditions \cite{6362517}. More importantly, in a highly automated environment, it is essential that the network itself can interpret, reason, and act upon situations that have not been explicitly encoded or trained for. This calls for a new generation of generalizable, adaptive, and autonomous management frameworks.

\begin{table*}[h]
\centering
\caption{Overview of Representative Digital Twin Approaches in 6G Literature}
\label{tab:related_work}
\renewcommand{\arraystretch}{1.2}
\scriptsize
\begin{tabular}{|p{1.5cm}|p{5.7cm}|p{6cm}|}
\hline
\textbf{Work} & \textbf{Scope and Contribution} & \textbf{Gap with Respect to Our Approach} \\
\hline
\cite{10597058} & Architectural vision for federated DT-based automation in 6G with AI-native orchestration. & Lacks semantic generalization or intent-based decision adaptation. \\
\hline
\cite{9491087} & DT-driven edge computing via adaptive placement and migration under dynamic channel states. & Focuses on latency minimization; no knowledge-based intent reasoning. \\
\hline
\cite{Li2025} & Broad survey of DT technologies and their applications in 6G systems. & Descriptive scope; does not propose an integrated management pipeline. \\
\hline
\cite{10283539} & Fully autonomous 6G with dual-loop orchestration using GNN and GAN-based modeling. & Lacks fine-grained semantic matching or zero-shot adaptability. \\
\hline
\cite{10107755} & SDN-enabled DT virtualization and DRL-based orchestration for service provisioning. & Action selection is tightly coupled to training; lacks abstraction and reuse. \\
\hline
\cite{10659965} & Urban-scale DT architecture for smart city services with distributed AI at the edge. & Targets large-scale sensing and optimization, not network control logic. \\
\hline
\cite{10975956} & Wireless DT for autonomous vehicles; real-time radio and mobility simulation. & Use-case specific; does not generalize to broader 6G management. \\
\hline
\end{tabular}
\end{table*}

In this paper, we propose a novel architectural framework designed to enhance the adaptability and autonomy of 6G network management systems. Our approach introduces a modular intelligence layer into the network management stack that observes evolving network conditions, interprets their contextual meaning, and selects appropriate management responses even for events that have never occurred before. At the core of this design lies a Digital Twin environment, which continuously mirrors the physical network’s behavior and enables contextual interpretation of network events through safe, real-time reflection. This twin-driven understanding is coupled with a generalization-capable decision layer that can infer appropriate actions without relying exclusively on labeled training data. The proposed framework is fully compatible with the Knowledge-Defined Networking (KDN) paradigm, supporting knowledge extraction, semantic representation, and reasoning across different stages of the network management cycle. The main contributions of this work are summarized as follows:
\begin{itemize}
  \item We propose a Digital Twin-assisted architecture that enables the network to observe and semantically interpret evolving behaviors in a virtual reflection of the real network.
  \item We introduce a generalization-based decision mechanism that can proactively determine suitable management actions even for previously unseen events, enhancing autonomous adaptability.
  \item We design a modular integration with the Knowledge Plane of KDN, enabling knowledge sharing and reuse by embedding semantic understanding into the management loop.
\end{itemize}

\section{Related Work}
Recent studies in the literature have extensively explored the integration of Digital Twin technology into 6G networks from various architectural and functional perspectives. In this section, we provide a structured analysis of representative works, following a progression from general architectural frameworks to more specific applications. One of the prominent architectural perspectives is presented in \cite{10597058}, which introduces the 6G-TWIN consortium’s vision for embedding Network Digital Twins within AI-native 6G architectures to address growing complexity in future networks. The proposed framework enables secure, federated closed-loop automation using digital replicas of the physical network for predictive planning and orchestration. Key components include harmonized data repositories, federated simulation engines, and AI-driven service management. The approach emphasizes infrastructure-level modeling and validation through sandbox environments. A DT-enhanced edge computing approach is discussed in \cite{9491087}, which proposes a wireless edge architecture that integrates Digital Twin (DT) models with edge computing for dynamic and latency-sensitive 6G scenarios. It formulates edge association as a dual-stage problem: DT placement and DT migration, and proposes deep reinforcement learning and transfer learning algorithms to optimize system latency and user utility. 

A broader perspective on the enabling technologies and evolution of DTs is provided in \cite{Li2025}, which presents a comprehensive exploration of digital twin (DT) technology and its pivotal role in shaping future 6G wireless communication networks. It reviews the evolution and key enabling technologies of DTs including modeling, data fusion, AI-driven analytics, and immersive visualization and outlines their practical applications in latency-sensitive edge computing, terahertz and RIS-assisted communication, aerial MEC, and non-terrestrial networks. A fully autonomous 6G framework based on DTs is introduced in \cite{10283539}, where the authors present a digital twin-based architecture to enable full autonomy in 6G wireless networks. The proposed framework combines real-time digital twin tracking with predictive planning through a dual-loop orchestration strategy. It facilitates decentralized, intent-driven control across wireless, transport, and core network segments using native-AI techniques. Key components include harmonized multi-source datasets, simulation-driven modeling (with GNNs and GANs), and AI-based service management for tasks such as beamforming optimization and dynamic RAN slicing. The approach emphasizes self-evolution and closed-loop automation enabled by digital replicas and situational awareness.

In \cite{10107755}, a software-defined 6G digital twin network architecture is proposed to enable adaptive service provisioning through digital twin function virtualization. Physical and virtual twin resources are decoupled into dynamic pools, allowing flexible orchestration. Proximal Policy Optimization, a deep reinforcement learning algorithm, is used to match virtualized twin functions to shifting service demands. The architecture incorporates SDN/OpenFlow-based collaboration among edge servers and optimizes parameters such as synchronization latency and operational cost. For urban-scale digital twin applications, \cite{10659965} explores how 6G networks can support smart city services by leveraging ultra-high data rates, low latency, and high-density connectivity for real-time urban modeling and service optimization. It introduces a triple-layered architecture integrating physical networks, digital twins, and computing networks, with native AI for distributed training and edge inference. The paper also discusses challenges such as data heterogeneity, system scalability, and privacy protection. A use-case-specific digital twin architecture for autonomous vehicles is introduced in \cite{10975956}, which proposes a wireless digital twin architecture to support 6G-connected autonomous vehicles by simulating urban environments and radio wave propagation in real time. It integrates traffic simulation, 3D modeling, and ray tracing to evaluate mobility-aware communication scenarios. The architecture addresses two key challenges: dynamic blockage prediction in non-line-of-sight conditions and multi-user hybrid beamforming for coordinating roadside units with moving vehicles.

As summarized in Table~\ref{tab:related_work}, existing studies address various aspects of Digital Twin integration, but none provide a generalizable, semantics-driven control architecture with zero-shot adaptability. Our work aims to bridge this gap through a unified, knowledge-infused framework.

\begin{figure*}[t]
  \centering
  \includegraphics[width=0.6\textwidth]{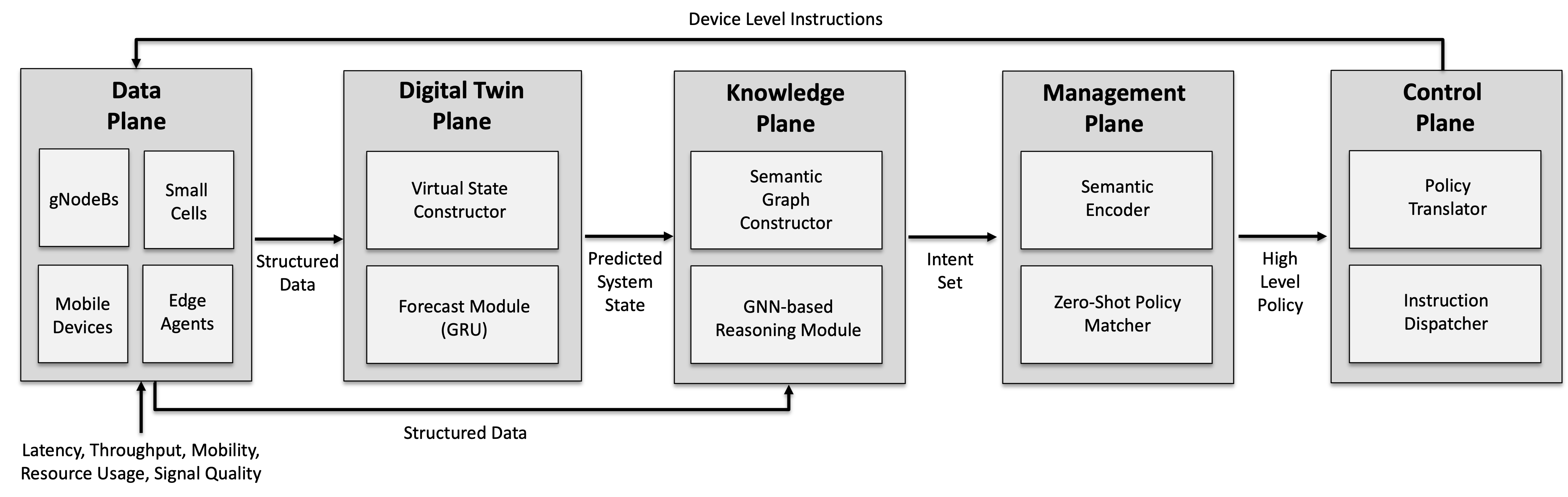}
  \caption{Layered architecture of the proposed KDN-based system with Digital Twin and semantic decision modules.}
  \label{fig:architecture}
\end{figure*}

\section{The Proposed Approach}
\subsection{System Overview}
The proposed architecture builds on the Knowledge-Defined Networking (KDN) paradigm, which introduces a multi-plane structure for enabling intelligent and autonomous network operations. As shown in Fig.~\ref{fig:architecture}, the system is organized into five core layers: the Data Plane, the Digital Twin Layer, the Knowledge Plane, the Management Plane, and the Control Plane. Each layer has a distinct role in sensing, modeling, interpreting, deciding, and acting upon network conditions. Our design introduces specialized modules into this architecture to support semantic understanding and generalizable decision-making, particularly for novel or unseen network events.

At the bottom of the stack, the \textbf{Data Plane} consists of physical and virtualized network infrastructure elements that generate real-time telemetry, traffic statistics, and performance indicators. These observations are collected and forwarded to the upper layers for processing and abstraction. The \textbf{Digital Twin Plane} acts as a live and predictive mirror of the physical network, maintaining a virtualized state that evolves with incoming telemetry. It captures both statistical metrics and structural configurations, and produces short-term state forecasts to anticipate future conditions. This layer enables safe simulation, temporal reasoning, and proactive insight generation without perturbing the operational environment. At the core of reasoning, the \textbf{Knowledge Plane} processes the virtual state generated by the Digital Twin. It constructs a semantic abstraction in the form of a graph representation, where nodes reflect functional components and edges encode interactions. A reasoning module embedded in this layer performs intent derivation, anomaly detection, and context interpretation based on the evolving semantic structure. The \textbf{Management Plane} creates the semantic representations and determines appropriate management actions, even in the absence of prior training for a specific event type. It performs semantic matching between current intents and pre-learned policy embeddings using zero-shot learning. The selected policy is then dispatched downstream. Finally, the \textbf{Control Plane} handles routing, resource allocation, and network reconfiguration tasks. It translates abstract policy decisions into low-level control actions and applies them to the Data Plane through programmable interfaces. This ensures fast and accurate enforcement of decisions made by the upper layers.

Overall, the proposed architecture maintains modular compatibility with KDN while extending it with context-awareness and zero-shot adaptability. The inclusion of a dedicated Digital Twin Layer enables reflective modeling and short-horizon prediction, which enrich the semantic reasoning and decision processes in the Knowledge and Management planes.

\subsection{Operational Flow and Functional Modules}
As explained above, the proposed architecture operates across five coordinated layers: the Data Plane, Control Plane, Knowledge Plane, Digital Twin Plane, and Management Plane. This layered structure enables an autonomous and context-aware decision-making pipeline for future 6G networks. The details of the proposed approach could be summarized as follows:

\subsubsection{Data Plane: Real-Time Observation and Collection}
This layer is responsible for real-time sensing and extraction of raw network parameters from the physical infrastructure, including gNodeBs, small cells, and mobile devices. At each time step $t$, the collected dataset is represented as a multidimensional vector set $
\mathcal{D}_t = \{d_t^{(1)}, d_t^{(2)}, \dots, d_t^{(n)}\}, \quad d_t^{(i)} = (\ell_t^{(i)}, \tau_t^{(i)}, \mu_t^{(i)}, \rho_t^{(i)}, \theta_t^{(i)})$. Here, $\ell_t^{(i)}$ is the end-to-end latency of user $i$ at time $t$. $\tau_t^{(i)}$ represents the instantaneous throughput. $\mu_t^{(i)}$ shows the user mobility level (e.g., handover frequency). $\rho_t^{(i)}$ is the local resource utilization (CPU/memory/bandwidth). Finally, $\theta_t^{(i)}$ corresponds to the error rate or signal quality (e.g., SINR or BER).

All collected metrics are normalized and packetized via edge agents embedded in RAN elements. These agents ensure real-time preprocessing (e.g., timestamping, compression, tagging), and forward the structured data $\mathcal{D}_t$ to the \textit{Digital Twin Engine} residing in the \textit{Knowledge Plane}. These preprocessing operations are performed by lightweight \textit{Edge Agents}, which include a modular \textit{Preprocessing Unit} responsible for timestamping, filtering, and formatting raw telemetry before transmission to the upper layers. This raw stream forms the basis for continuous model updates, inference, and predictive analytics in higher layers.

\subsubsection{Digital Twin Layer: Virtual State Construction and Predictive Modeling}
This is the central engine of the Knowledge Plane, responsible for maintaining an up-to-date and semantically rich virtual representation of the physical network environment. The layer is composed of two core submodules: a \textit{Virtual State Constructor} that maintains the current virtual topology and statistics, and a \textit{Short-Term Forecaster} based on a GRU model that anticipates near-future states \cite{10001102}. Upon receiving incoming data streams $\mathcal{D}_t$ from the Data Plane, it synthesizes this information into the virtual state space $\hat{\mathcal{S}}_t$, covering both structural and statistical characteristics of the network. Accordingly, the virtual state is modeled as given in Eq. \ref{e1}. In Eq. \ref{e1}, $\phi_\text{stat}$ captures time-varying aggregate metrics (e.g., queue lengths, link utilization), while $\phi_\text{struct}$ encodes the topological configuration, including node states and connectivity dynamics.

\begin{equation}\label{e1}
    \hat{\mathcal{S}}_t = \phi_\text{stat}(\mathcal{D}_t) \cup \phi_\text{struct}(\mathcal{D}_t)
\end{equation}

Beyond real-time mirroring, the layer supports short-horizon forecasting via an internal predictive model $f_{\text{DT}}$, which learns temporal patterns from a sliding window of past observations as given in Eq. \ref{e2}.

\begin{equation}\label{e2}
    \hat{\mathcal{S}}_{t+\Delta} = f_{\text{DT}}(\mathcal{D}_{t-k:t})
\end{equation}

In Eq. \ref{e2}, $k$ denotes the number of past time steps used for prediction, and $\Delta$ specifies the look-ahead duration. The model $f_{\text{DT}}$ is implemented as a lightweight Gated Recurrent Unit (GRU)-based neural predictor, chosen for its balance between temporal modeling accuracy and computational efficiency in real-time environments.

The forecasted state $\hat{\mathcal{S}}_{t+\Delta}$ is not merely supplementary but constitutes a core enabler of the Knowledge Plane’s proactive intelligence. It supports early identification of congestion build-up, SLA violation trajectories, and dynamic shifts in network conditions. Thus empowering mitigation decisions before service degradation becomes observable.

\subsubsection{Knowledge Plane: Abstraction, Reasoning, and Semantic Mapping}
The predicted system state $\hat{\mathcal{S}}_{t+\Delta}$, obtained from the Digital Twin Layer, is abstracted into a dynamic knowledge graph $\mathcal{G}_{t+\Delta} = (\mathcal{V}, \mathcal{E}, \mathbf{X})$. Each node $v_i \in \mathcal{V}$ represents a network element (e.g., gNB, user slice, or control function), and each edge $(v_i, v_j) \in \mathcal{E}$ encodes a context-specific relation such as traffic dependency, orchestration hierarchy, or control signaling. The node feature matrix $\mathbf{X}$ captures semantically enriched representations derived from historical measurements, forecasted metrics, and SLA-related indicators.

These abstractions are generated by a dedicated \textit{Semantic Graph Constructor}, which transforms predicted states into graph structures. A \textit{GNN-Based Reasoning Module} then operates on this graph to infer causality, detect latent anomalies, and generate intents. This reasoning process is embedded in a continuous graph update cycle, where the knowledge graph is periodically refreshed using both real-time measurements and the look-ahead state $\hat{\mathcal{S}}_{t+\Delta}$ generated by the Digital Twin Layer. This forward-aware graph evolution ensures that the reasoning mechanism not only reflects the current system state but also anticipates near-future dynamics and causality.

As a result of each graph update cycle, a candidate intent set $\mathcal{I}_{t+\Delta} = \{i_1, i_2, \dots\}$ is generated, where each $i_k$ represents a high-level proactive control directive such as rerouting, load balancing, or function migration. These intents are passed to the Management Plane for prioritization, arbitration, and distributed execution.

\subsubsection{Management Plane: Zero-Shot Policy Mapping and Decision Execution}
Upon receiving the candidate intent set $\mathcal{I}_t = \{i_1, i_2, \dots\}$ from the Knowledge Plane, the Management Plane is responsible for selecting and executing the most appropriate control actions. This plane is composed of two submodules: a \textit{Semantic Encoder}, which projects high-level intents into an embedding space, and a \textit{ZSL Policy Matcher}, which identifies the most semantically aligned policy vector from a pre-learned library. Rather than relying on retraining or explicit supervision for every unseen scenario, the Management Plane utilizes a Zero-Shot Learning (ZSL) mechanism to perform semantic matching between each intent and a set of pre-learned policy embeddings.

Each intent $i_j \in \mathcal{I}_t$ is first encoded into a latent vector representation $\phi(i_j)$ using the shared semantic encoder trained over a curated set of prior intent-policy pairs. This representation captures high-level operational goals, contextual constraints, and expected system impacts. The encoded intent is then compared to a stored policy embedding space $\Phi = \{\phi_1, \phi_2, \dots, \phi_m\}$ using cosine similarity as given in Eq. \ref{e3}. In Eq. \ref{e3}, $\text{sim}(\cdot)$ denotes the cosine similarity between embeddings. The selected policy $\phi^*(i_j)$ corresponds to an executable management directive such as rerouting, dynamic scaling, access restriction, or function migration. This decision is subsequently transmitted to the Control Plane for real-time enforcement. 

\begin{equation} \label{e3}
  \phi^*(i_j) = \arg\max_{\phi_k \in \Phi} \ \text{sim}(\phi(i_j), \phi_k) 
\end{equation}

The ZSL-based policy selection pipeline enables the system to respond to previously unseen situations by leveraging abstract semantic similarities, thereby enhancing adaptability, reducing reaction latency, and minimizing the need for retraining in dynamic 6G environments.

\subsubsection{Control Plane: Policy Translation and Enforcement}
Once a high-level policy $\phi^*(i_j)$ is selected by the Management Plane, the Control Plane is tasked with translating this abstract intent into concrete, low-level network actions. Internally, the Control Plane contains a \textit{Policy Translator} that maps semantic policies to executable commands, and an \textit{Instruction Dispatcher} that pushes these actions to the appropriate Data Plane elements. This includes generating flow table entries, updating scheduling parameters, reconfiguring slice templates, or invoking interface-level APIs, depending on the policy type and targeted infrastructure.

A policy translation module decomposes each semantic policy into one or more device-level instructions, which are then propagated to the corresponding Data Plane components (e.g., RAN nodes, edge routers, core orchestrators). To maintain responsiveness, this translation and dissemination process is designed to operate with minimal latency, avoiding centralized recomputation or policy re-synthesis.

By ensuring fast and accurate enforcement, the Control Plane finalizes the intent lifecycle initiated by the Knowledge and Management layers. The combined use of predictive digital twin states, semantic reasoning, and zero-shot policy matching enables the entire pipeline to react effectively to previously unseen scenarios while preserving low control overhead. This is also a key requirement for scalable and autonomous 6G systems.

Overall, to ensure continuous learning and adaptation, the proposed architecture incorporates a closed-loop feedback mechanism that connects the Control Plane back to the upper layers. Once a policy is enforced by the Control Plane, its operational impact is collected in real time by the Data Plane and forwarded upstream. This feedback is first processed by the Digital Twin Layer, which integrates post-action system responses into its virtual model, refining future predictions and correcting any discrepancies between estimated and observed outcomes. Simultaneously, the Knowledge Plane updates the knowledge graph $\mathcal{G}_t$ with new observations, enabling it to detect persistent anomalies, re-learn dependency structures, and generate more informed intents in the next cycle. The Management Plane, in turn, uses this feedback to update the semantic encoder's representation space, gradually enhancing the generalization capability of the Zero-Shot Learning module through implicit experience accumulation. Through this closed-loop operation, the system forms a self-adaptive control cycle that improves over time without requiring extensive retraining or manual intervention.

\section{Performance Evaluation}
\subsection{Simulation Environment and Details}
To evaluate the effectiveness of the proposed framework, we implemented a realistic simulation environment that emulates ultra-dense urban scenarios with heterogeneous infrastructure and mobile users. The physical network is modeled as a dynamic graph, where vertices include multiple gNodeBs (macro and small cells) and user equipment (UEs), and edges represent active communication links. Network parameters such as latency, throughput, handover frequency, and resource utilization are synthetically generated based on stochastic distributions reflecting realistic 6G traffic and mobility patterns. Key aspects of the simulation setup include:
\begin{itemize}
  \item {Network Topology:} 12 heterogeneous base stations with overlapping coverage; UE density varies between 300 and 600 to simulate peak and off-peak load.
  \item {Mobility Model:} Gauss-Markov model applied to user movements to produce correlated velocity and direction changes.
  \item {Data Collection Interval:} Network telemetry is collected every 100 milliseconds and forwarded to upper layers for real-time processing.
  \item {Digital Twin Configuration:} The Digital Twin Layer uses a GRU-based predictor trained to forecast network state over a look-ahead horizon $\Delta = 3$ time steps, with historical context window size $k = 5$.
  \item {Knowledge Graph Updating:} The Knowledge Plane refreshes the semantic graph at every time step using both actual and predicted data from the Digital Twin Layer.
  \item {Zero-Shot Learning Setup:} The Management Plane’s policy matcher is trained on 15 distinct network management intents, each embedded in a 64-dimensional semantic space to facilitate generalization to unseen scenarios.
\end{itemize}

We evaluate the proposed approach against following two widely adopted network management approaches:

\begin{itemize}
    \item {Static Policy Control:} A conventional rule-based system in which network policies are predefined based on prior knowledge. Management actions are triggered when observed events match known patterns.
    \item {Supervised ML with Retraining:} A learning-based architecture that uses a deep neural network trained on labeled network event logs. The model classifies network conditions and suggests actions, with periodic retraining to maintain accuracy over time.    
\end{itemize}

\begin{figure*}[h]	
	\centering		
        \subfloat[]{%
		\includegraphics[width=0.3\textwidth]{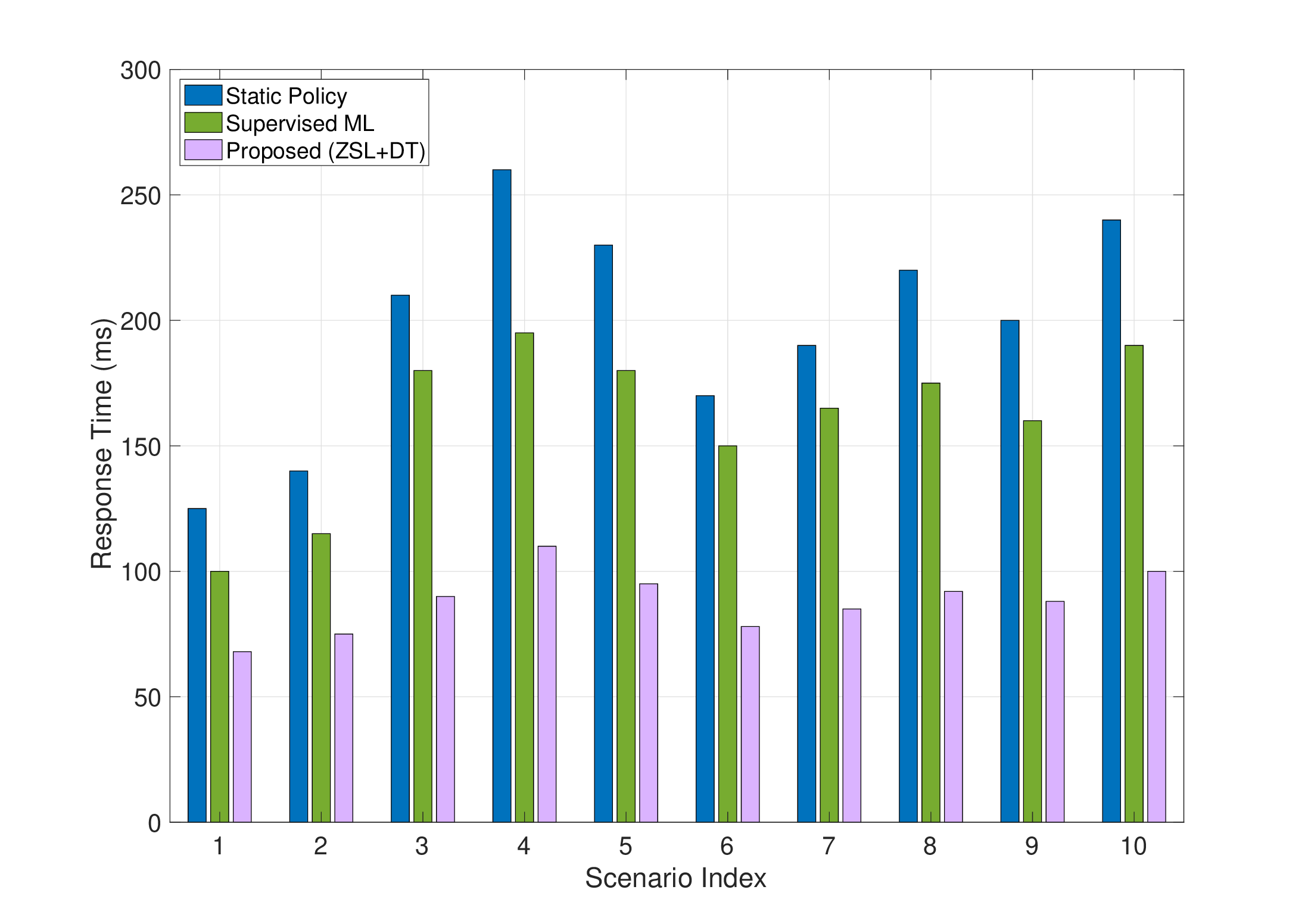}%
		\label{g1}%
	} 
     \subfloat[]{%
		\includegraphics[width=0.3\textwidth]{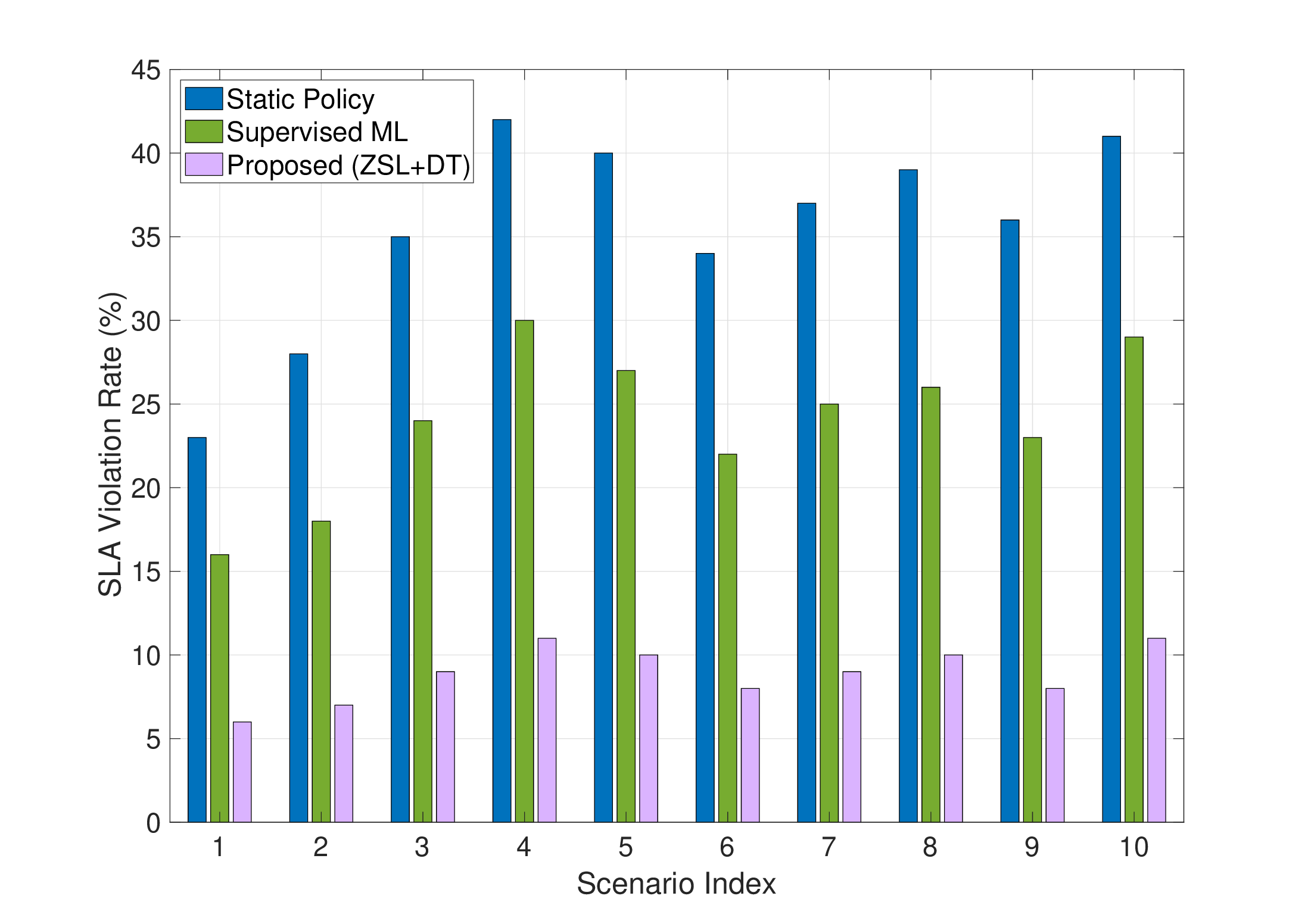}%

		\label{g2}%
	} %
         \subfloat[]{%
		\includegraphics[width=0.3\textwidth]{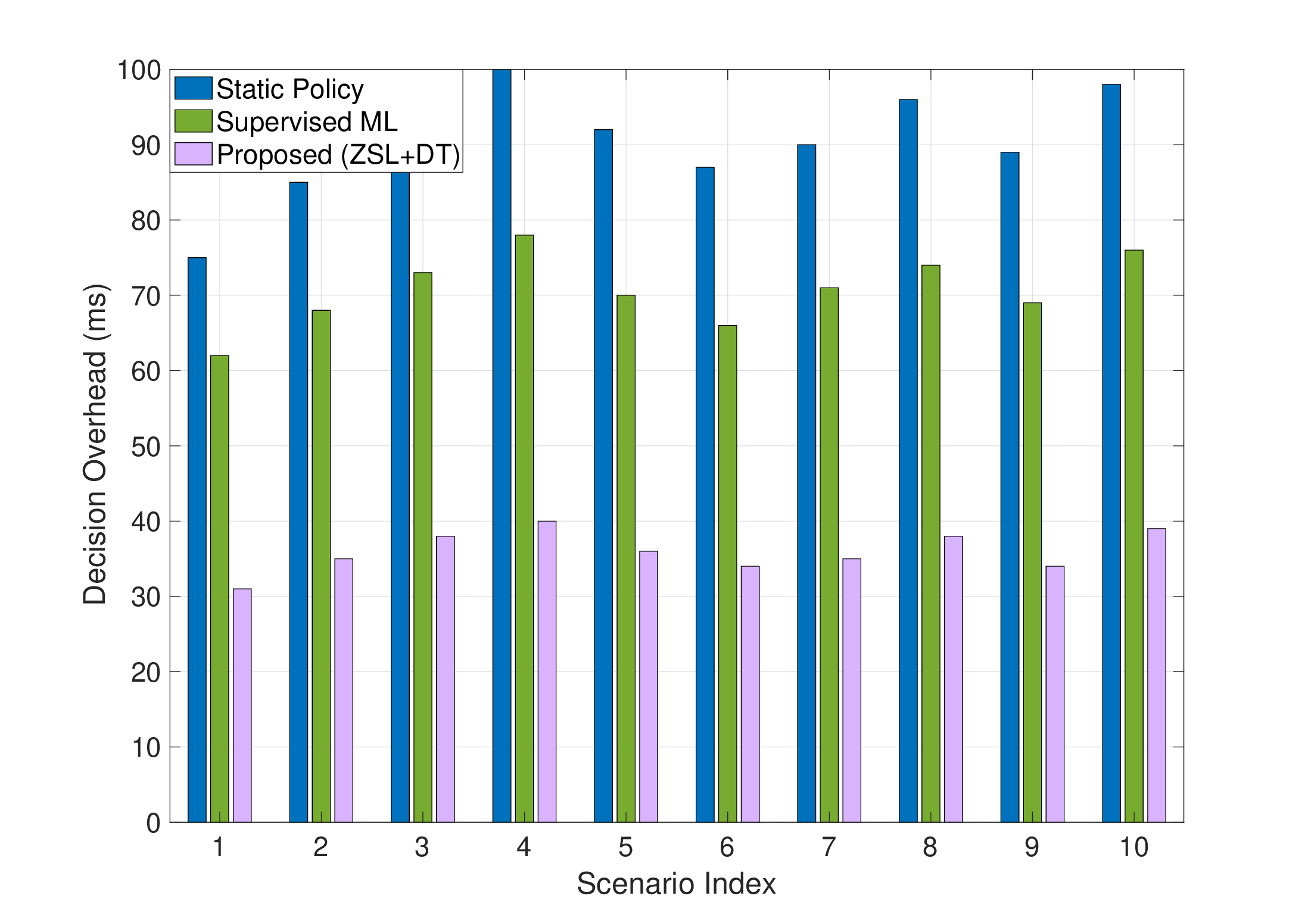}%
		\label{g3}%
	} %
	\caption{Evaluations for Response Time, SLA Violation Rate, and Decision Overhead}
	\label{r6}
\end{figure*} 

The proposed framework, denoted as ZSL-DT (Ours), combines predictive modeling via a Digital Twin with Zero-Shot Learning to infer suitable policies for novel network states, without requiring labeled training data for every scenario. Also, to ensure a comprehensive evaluation, we define ten diverse network scenarios covering various operational contexts such as user mobility, traffic surges, resource constraints, and fault conditions. These scenarios are summarized in Table~\ref{tab:scenarios} and serve as the basis for comparative analysis across all performance metrics.

\begin{table}[ht]
\centering
\caption{Scenario Index and Descriptions}
\label{tab:scenarios}
\scriptsize
\begin{tabular}{|c|p{5cm}|}
\hline
\textbf{Scenario} & \textbf{Description} \\ \hline \hline
1 & Light traffic, static users, rural macro-cell. \\\hline
2 & Moderate load with low-mobility users. \\\hline
3 & Sudden event-driven traffic burst. \\\hline
4 & Edge congestion under multi-slice setup. \\\hline
5 & High-mobility users causing handover bursts. \\\hline
6 & Urban interference with co-channel users. \\\hline
7 & Link degradation due to external factors. \\\hline
8 & Delays from service chaining overhead. \\\hline
9 & SLA risks at resource-constrained edge. \\\hline
10 & Compound anomaly: burst + backhaul issue. \\\hline
\end{tabular}
\end{table}

\subsection{Performance Comparison}
\paragraph{Response Time}
Response time is defined as the delay between the occurrence of a network anomaly and the execution of the corresponding management action. As illustrated in Fig.~\ref{g1}, our system demonstrates a clear advantage in response latency. In Static Policy Control, responses are instantaneous for known conditions but fail to generalize beyond predefined triggers, leading to missed detections. The Supervised ML model requires inference plus retraining time when encountering unseen events, resulting in delayed adaptation. In contrast, our proposed ZSL-DT framework leverages the proactive forecasting capability of the Digital Twin and semantic matching via Zero-Shot Learning, enabling sub-second response times even for novel scenarios. The elimination of retraining cycles and simulation-assisted anticipation significantly reduce decision latency.

\paragraph{SLA Violation Rate}
This metric quantifies the percentage of user sessions that experience service degradation beyond agreed thresholds (e.g., excessive latency or packet loss). As shown in Fig.~\ref{g2}, the proposed system significantly outperforms baseline approaches in minimizing SLA violations. Static Policy Control lacks contextual understanding and fails to detect early degradation trends, leading to higher SLA violations, especially under unforeseen conditions. The Supervised ML model performs better but is sensitive to training distribution shifts and cannot preempt violations without prior examples. Our ZSL-DT architecture predicts future system states through the Digital Twin, allowing early detection of SLA degradation patterns. It maps abstract intents to actionable policies even in the absence of historical labels, resulting in a significantly reduced SLA violation rate across mobility and load spikes.

\paragraph{Decision Overhead}
Decision overhead measures the computational and communication resources consumed during policy inference and enforcement. Fig.~\ref{g3} summarizes the comparative decision overhead observed during simulations. Static methods incur negligible overhead due to simple rule execution but offer limited adaptability. Supervised ML models introduce moderate to high overhead depending on the depth of the network and retraining frequency, especially when deployed at edge nodes. Our ZSL-DT framework maintains a lightweight GRU-based forecaster and low-dimensional semantic embedding space, ensuring scalable performance. 

\section{Conclusion}
We presented a modular 6G management architecture that combines Digital Twin modeling, semantic reasoning, and zero-shot learning to enable proactive and generalizable decision-making. By leveraging real-time virtual replicas and semantic policy matching, the system can handle unseen scenarios without retraining. Simulations show notable improvements in response time, SLA compliance, and decision overhead.

\bibliographystyle{IEEEtran}
\bibliography{ref}

\end{document}